\documentclass[twocolumn,showpacs,amsmath]{revtex4}%
\usepackage{amsfonts}
\usepackage{amsmath}
\usepackage{amssymb}
\usepackage{graphicx}
\usepackage{color}%
\providecommand{\U}[1]{\protect\rule{.1in}{.1in}}
\definecolor{darkgreen}{rgb}{0,.5,0}

\begin{document}
\title[ ]{Multiphoton coherent population oscillation}
\author{A. V. Sharypov}
\affiliation{Kirensky Institute of Physics, 50 Akademgorodok, Krasnoyarsk, 660036, Russia}
\affiliation{Siberian Federal University, 79 Svobodny Ave., Krasnoyarsk, 660041, Russia}
\author{A. D. Wilson-Gordon }
\affiliation{Department of Chemistry, Bar-Ilan University, Ramat Gan 52900, Israel}


\begin{abstract}
We study the bichromatic driving of a two-level system which displays
long-lived coherent population oscillations (CPO). We show that under certain
conditions, multiphoton parametric interaction leads to the appearance of CPO
resonances at the subharmonic frequencies. In addition, in the region of the
CPO resonances, there is strong parametric interaction between the weak
sideband components of the electromagnetic field.

\end{abstract}
\maketitle

\section{\label{sec:level1}Introduction}

Non-linear interaction between electromagnetic fields can lead to the
appearance of resonances with a bandwidth which is much narrower than the
unperturbed natural linewidth. The most familiar effects that can give such
narrowband response are electromagnetically induced transparency (EIT)
\cite{EITrev} and coherent population oscillations (CPO) \cite{CPO}. In the
EIT\ and CPO cases, the strong pump can form a transparency window in the
probe absorption spectrum accompanied by steep dispersion of the refractive
index that leads to effects such as slow light\emph{ }%
\cite{HarrisPRA1992,BoydSL}. The narrowband medium response can also appear in
the four-wave mixing (FWM) process \cite{FriedmannCPO}, leading under certain
conditions to narrowband biphoton generation due to EIT \cite{NBHarris,NB} or
CPO \cite{NBCPO}. Although there are similarities between EIT and CPO, these
effects are actually different in nature, and the properties of the non-linear
response in each case are determined by different system parameters. For
example, in the case of EIT, the minimal bandwidth of the resonance is
determined by the transverse relaxation rate of the two-photon transition,
whereas in the case of CPO, the minimal bandwidth is determined by the
effective longitudinal relaxation rate; in the EIT case, the resonance appears
at zero two-photon detuning, whereas in the CPO case, the non-linear resonance
is centered at the pump frequency. The fact that in the CPO case, the weak
signal is always centered at the pump frequency makes it difficult to filter
out the weak signal from the strong pump and incoherent scattering.

Here, we analyze the response of a two-level system (TLS) that displays
long-lived coherent population oscillations in the presence of a bichromatic
pump, and one or two weak fields. The two sidebands of the pump are
symmetrically displaced from the pump frequency $\omega_{0}$\thinspace\ and in
the case where the scanning is realized by two probe fields, these fields are
also symmetrically displaced from $\omega_{0}$ (see Fig. \ref{IntScheme}). The
interaction of the TLS with a polychromatic field has been studied from many
different aspects; see, for example, \cite{FicekProgrOpt} and references
therein. In the bichromatic fields, resonances at the subharmonic frequencies
$\omega_{sub}=\omega_{0}\pm\left(  2n+1\right)  \delta$
\cite{Zhu1990,Ficek1993} appear where $\delta$ is the frequency difference
between the two pumps and $n$\ is an integer. Here we demonstrate that these
resonances can appear under CPO conditions with a width determined by the
effective longitudinal relaxation rate. We also demonstrate that there is
effective parametric interaction between two weak probes tuned in the region
of the two symmetrically displaced CPO resonances. Under certain conditions
this parametric interaction can appear at a pump Rabi frequency well below the
transverse relaxation rate of the TLS.

\section{\label{sec:level2}The model}

\begin{figure}
[ptb]
\begin{center}
\includegraphics[
height=2.9603in,
width=2.6662in
]%
{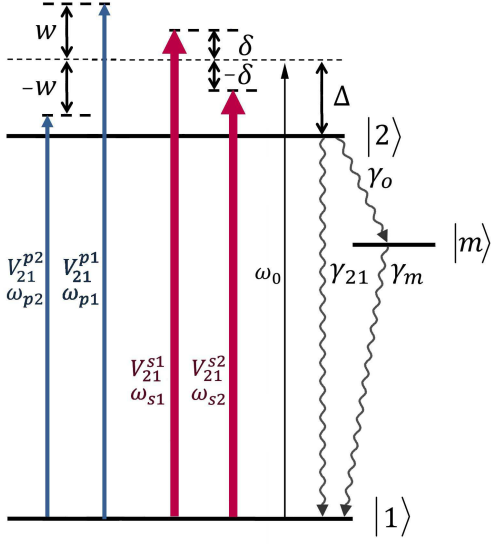}%
\caption{A two-level system $\left\vert 2\right\rangle -\left\vert
1\right\rangle $ interacting with bichromatic pump field $V_{21}^{s1,s2}$ and
bichromatic probe field $V_{21}^{p1,p2}$. $\gamma_{21}$ longitudinal
relaxation from the excited state to the ground, $\gamma_{o}$ - longitudinal
relaxation rate from the excited state to the intermediate metastable state
$\left\vert m\right\rangle $ and $\gamma_{m}$ is the longitudinal relaxation
rate from intermediate state to the ground. }%
\label{IntScheme}%
\end{center}
\end{figure}

Let us consider the two-level quantum system composed of levels $\left\vert
1\right\rangle $ and $\left\vert 2\right\rangle $ interacting with the
electromagnetic field $E$. It is assumed that system has an additional
off-resonant quantum metastable state $\left\vert m\right\rangle $ which is
radiatively coupled to the TLS (see Fig. \ref{IntScheme}).%
%TCIMACRO{\FRAME{ftbpFU}{2.6662in}{2.9603in}{0pt}{\Qcb{A two-level system
%$\left\vert 2\right\rangle -\left\vert 1\right\rangle $ interacting with
%bichromatic pump field $V_{21}^{s1,s2}$ and bichromatic probe field
%$V_{21}^{p1,p2}$. $\gamma_{21}$ longitudinal relaxation from the excited state
%to the ground, $\gamma_{o}$ - longitudinal relaxation rate from the excited
%state to the intermediate metastable state $\left\vert m\right\rangle $ and
%$\gamma_{m}$ is the longitudinal relaxation rate from intermediate state to
%the ground. }}{\Qlb{IntScheme}}{interactionscheme4a.eps}%
%{\special{ language "Scientific Word";  type "GRAPHIC";
%maintain-aspect-ratio TRUE;  display "USEDEF";  valid_file "F";
%width 2.6662in;  height 2.9603in;  depth 0pt;  original-width 5.2762in;
%original-height 5.8617in;  cropleft "0";  croptop "1";  cropright "1";
%cropbottom "0";  filename 'InteractionScheme4a.eps';file-properties "XNPEU";}}
%}%
%BeginExpansion
%EndExpansion
The set of Bloch equations for the present system has the following form:%
\begin{equation}
\left(  \frac{d}{dt}+\Gamma_{21}+i\omega_{21}\right)  \rho_{21}+iV_{21}\left(
\rho_{11}-\rho_{22}\right)  =0 \label{Bloch_ro21}%
\end{equation}%
\begin{equation}
\left(  \frac{d}{dt}+\Gamma_{21}-i\omega_{21}\right)  \rho_{12}-iV_{12}\left(
\rho_{11}-\rho_{22}\right)  =0 \label{Bloch_ro12}%
\end{equation}%
\begin{equation}
V_{12}\rho_{21}-iV_{21}\rho_{12}+\frac{d\rho_{11}}{dt}-\gamma_{21}\rho
_{22}-\gamma_{m}\rho_{m}=0 \label{Bloch_ro11}%
\end{equation}%
\begin{equation}
\frac{d\rho_{m}}{dt}+\gamma_{m}\rho_{m}-\gamma_{o}\rho_{22}=0
\label{Bloch_ro33_0}%
\end{equation}
where $V_{21}$ is the total Rabi frequency, and $\Gamma_{21}$ is the
transverse relaxation rate of the $\left\vert 2\right\rangle -\left\vert
1\right\rangle $ transition, and $\gamma_{21}$, $\gamma_{o}$, and $\gamma_{m}$
are the population relaxation constants from the excited state to the ground
state, from the excited state to the metastable state, and from the metastable
to the ground state, respectively. Further, we assume that the system is
closed and that the total population in the system is conserved so that%
\begin{equation}
\rho_{11}+\rho_{22}+\rho_{m}=1. \label{Conservation}%
\end{equation}
Let us introduce the population inversion
\begin{equation}
r=\rho_{22}-\rho_{11}, \label{r_Def}%
\end{equation}
and, using Eqs. (\ref{Conservation}) and (\ref{r_Def}), define $\rho_{11}$ and
$\rho_{22}$ in terms of $r$ and $\rho_{m}$. We then rewrite Eqs.
(\ref{Bloch_ro21}) - (\ref{Bloch_ro33_0}) and obtain the following set of
equations:
\begin{equation}
\left(  \frac{d}{dt}+\Gamma_{21}+i\omega_{21}\right)  \rho_{21}-iV_{21}r=0,
\label{Bloch_ro21_mod1}%
\end{equation}%
\begin{equation}
\left(  \frac{d}{dt}+\Gamma_{21}-i\omega_{21}\right)  \rho_{12}+iV_{12}r=0,
\label{Bloch_ro12_mod}%
\end{equation}%
\begin{align}
&  2iV_{12}\rho_{21}-2iV_{21}\rho_{12}-\left(  \frac{dr}{dt}+\gamma_{21}%
+\frac{\gamma_{o}}{2}\right)  r\nonumber\\
&  +\left(  \gamma_{21}-\gamma_{m}+\frac{\gamma_{o}}{2}\right)  \rho
_{m}=\gamma_{21}+\frac{\gamma_{o}}{2}, \label{Bloch_ro_r_mod}%
\end{align}%
\begin{equation}
-\frac{\gamma_{o}}{2}r+\left(  \frac{d}{dt}+\gamma_{m}+\frac{\gamma_{o}}%
{2}\right)  \rho_{m}=\frac{\gamma_{o}}{2}. \label{Bloch_ro_m_mod}%
\end{equation}

\subsection{The electromagnetic field}

We now specify the electromagnetic fields acting on the system. We consider
the case of the bichromatic pumping of the TLS at frequencies $\omega_{s1}$
and $\omega_{s2}$, and define the detuning of the pump fields from resonance
in the following way%
\begin{equation}
\Delta=\omega_{0}-\omega_{21}, \label{det_w1}%
\end{equation}
where $\omega_{0}=({\omega_{s1}+\omega_{s2}})/{2}$ and $\omega_{21}$ is the
resonance frequency of the $\left\vert 1\right\rangle \rightarrow\left\vert
2\right\rangle $ transition. We also apply a weak probe field which is a
combination of two sidebands arranged symmetrically with respect to
$\omega_{0}$\ with frequencies $\omega_{p1}$ and $\omega_{p2}$. The Rabi
frequency for the total electromagnetic field can be written in the form%
\begin{equation}
V_{21}=\left(  V_{21}^{s1}e^{-i\delta t}+V_{21}^{s2}e^{i\delta t}+V_{21}%
^{p1}e^{-iwt}+V_{21}^{p2}e^{iwt}\right)  e^{-i\omega_{0}t}, \label{V21}%
\end{equation}
where $V_{21}^{s1}$, $V_{21}^{s2}$, $V_{21}^{p1}$, and $V_{21}^{p2}$ are the
Rabi frequencies of the corresponding fields. We also define the following
detunings:%
\begin{equation}
\delta=\omega_{s1}-\omega_{0}=-\left(  \omega_{s2}-\omega_{0}\right)  \text{,
}w=\omega_{p1}-\omega_{0}=-\left(  \omega_{p2}-\omega_{0}\right)  .
\label{freq_notation}%
\end{equation}
Next we move to the frame rotating at frequency $\omega_{0}$ and define%
\begin{equation}
\sigma_{21}=\rho_{21}e^{i\omega_{0}t},~\sigma_{12}=\rho_{12}e^{-i\omega_{0}%
t},~\sigma_{0}=r,~\sigma_{m}=\rho_{mm}. \label{ro_rep}%
\end{equation}
Substituting Eqs. (\ref{V21}) and (\ref{ro_rep}) into Eqs.
(\ref{Bloch_ro21_mod1})-(\ref{Bloch_ro_m_mod}), we obtain a set of equations
which we write in the matrix form%
\begin{equation}
BS=F, \label{BlochInMatrix2}%
\end{equation}
where%
\begin{equation}
S=\left(
\begin{array}
[c]{c}%
\sigma_{21}\\
\sigma_{12}\\
\sigma_{0}\\
\sigma_{m}%
\end{array}
\right)  \text{, }F=\left(
\begin{array}
[c]{c}%
0\\
0\\
-\left(  \gamma_{21}+\frac{\gamma_{o}}{2}\right) \\
\frac{\gamma_{o}}{2}%
\end{array}
\right)  , \label{S_matrix_def}%
\end{equation}
and the matrix $B$ is written as the sum of components oscillating at the
different frequencies%
\begin{equation}
B=O+D^{+\delta}e^{i\delta t}+D^{-\delta}e^{-i\delta t}+W^{+w}e^{iwt}%
+W^{-w}e^{-iwt}, \label{B_separation}%
\end{equation}
where the matrices are defined as%
\begin{align}
&  O=\frac{d}{dt}I\nonumber\\
&  +\left(
\begin{array}
[c]{cccc}%
\Gamma_{21}-i\Delta & 0 & 0 & 0\\
0 & \Gamma_{21}+i\Delta & 0 & 0\\
0 & 0 & \gamma_{21}+\frac{\gamma_{o}}{2} & \gamma_{m}-\frac{\gamma_{o}}%
{2}-\gamma_{21}\\
0 & 0 & -\frac{\gamma_{o}}{2} & \gamma_{m}+\frac{\gamma_{o}}{2}%
\end{array}
\right)  , \label{O_matrix_def}%
\end{align}%
\begin{align}
D^{-\delta}  &  =i\left(
\begin{array}
[c]{cccc}%
0 & 0 & -V_{21}^{~s1} & 0\\
0 & 0 & V_{12}^{~s2} & 0\\
-2V_{12}^{~s2} & 2V_{21}^{~s1} & 0 & 0\\
0 & 0 & 0 & 0
\end{array}
\right)  ,\text{ }\label{D_matrix_def}\\
D^{+\delta}  &  =i\left(
\begin{array}
[c]{cccc}%
0 & 0 & -V_{21}^{~s2} & 0\\
0 & 0 & V_{12}^{~s1} & 0\\
-2V_{12}^{~s1} & 2V_{21}^{~s2} & 0 & 0\\
0 & 0 & 0 & 0
\end{array}
\right)  ,\\
W^{-w}  &  =i\left(
\begin{array}
[c]{cccc}%
0 & 0 & -V_{21}^{p1} & 0\\
0 & 0 & V_{12}^{p2} & 0\\
-2V_{12}^{p2} & 2V_{21}^{p1} & 0 & 0\\
0 & 0 & 0 & 0
\end{array}
\right)  ,\label{W_matrix_def}\\
W^{+w}  &  =i\left(
\begin{array}
[c]{cccc}%
0 & 0 & -V_{21}^{p2} & 0\\
0 & 0 & V_{12}^{p1} & 0\\
-2V_{12}^{p1} & 2V_{21}^{p2} & 0 & 0\\
0 & 0 & 0 & 0
\end{array}
\right)  ,
\end{align}
and $I$ is the identity matrix of order 4.

\subsection{Expansion of the solution}

As can be seen from Eq. (\ref{B_separation}), the Hamiltonian of the system
has a periodic time dependence with two characteristic frequencies $\delta$
and $w$. In order to find a solution, we use the Floquet theorem and make a
harmonic expansion of the vector of the density-matrix elements $S$%
\begin{equation}
S=%
%TCIMACRO{\dsum \limits_{n=-\infty}^{+\infty}}%
%BeginExpansion
{\displaystyle\sum\limits_{n=-\infty}^{+\infty}}
%EndExpansion%
%TCIMACRO{\dsum \limits_{m=-\infty}^{+\infty}}%
%BeginExpansion
{\displaystyle\sum\limits_{m=-\infty}^{+\infty}}
%EndExpansion
S^{n,m}e^{i\left(  n\delta+mw\right)  t}. \label{S_decomposition}%
\end{equation}
After substituting Eq. (\ref{S_decomposition}) into Eq. (\ref{BlochInMatrix2}%
), we take the time derivative of the oscillating term and separate the
equations with different time dependence to obtain the matrix recurrence
relation%
\begin{align}
&  \left(  O+T\right)  S^{n,m}+D^{-\delta}S^{n+1,m}+D^{+\delta}S^{n-1,m}%
+\nonumber\\
&  W^{-w}S^{n,m+1}+W^{+w}S^{n,m-1}=F\delta_{n=0,m=0}, \label{mn_equation}%
\end{align}
where%
\begin{equation}
T=i\left(  n\delta+mw\right)  I.
\end{equation}

\subsection{Steady state approximation}

We now make the steady-state approximation and rewrite the matrix equation of
Eq. (\ref{mn_equation})\ as a set of four algebraic equations:%
\begin{align}
&  \sigma_{21}^{n,m}=iP_{21}^{n,m}\nonumber\\
&  \left(  V_{21}^{s1}\sigma_{0}^{n+1,m}+V_{21}^{s2}\sigma_{0}^{n-1,m}%
+V_{21}^{p1}\sigma_{0}^{n,m+1}+V_{21}^{p2}\sigma_{0}^{n,m-1}\right)  ,
\label{sigma21mn2}%
\end{align}%
\begin{align}
&  \sigma_{12}^{n,m}=-iP_{12}^{n,m}\nonumber\\
&  \left(  V_{12}^{s2}\sigma_{0}^{n+1,m}+V_{12}^{s1}\sigma_{0}^{n-1,m}%
+V_{12}^{p1}\sigma_{0}^{n,m-1}+V_{12}^{p2}\sigma_{0}^{n,m+1}\right)  ,
\label{sigma12mn2}%
\end{align}%
\begin{align}
&  \left[  \gamma_{21}+\frac{\gamma_{o}}{2}+i\left(  n\delta+mw\right)
\right]  \sigma_{0}^{n,m}-\left(  \gamma_{21}-\gamma_{m}+\frac{\gamma_{o}}%
{2}\right)  \sigma_{m}^{n,m}\nonumber\\
&  2iV_{21}^{s1}\sigma_{12}^{n+1,m}-2iV_{12}^{s1}\sigma_{21}^{n-1,m}%
-2iV_{12}^{s2}\sigma_{21}^{n+1,m}+\nonumber\\
&  2iV_{21}^{s2}\sigma_{12}^{n-1,m}+2iV_{21}^{p1}\sigma_{12}^{n,m+1}%
-2iV_{12}^{p1}\sigma_{21}^{n,m-1}+\nonumber\\
&  2iV_{21}^{p2}\sigma_{12}^{n,m-1}-2iV_{12}^{p2}\sigma_{21}^{n,m+1}=-\left(
\gamma_{21}+\frac{\gamma_{o}}{2}\right)  \delta_{n=0,m=0}, \label{sigma0mn}%
\end{align}%
\begin{equation}
\sigma_{m}^{n,m}=P_{m}^{n,m}\left(  \delta_{n=0,m=0}+\sigma_{0}^{n,m}\right)
, \label{sigma_m_mn}%
\end{equation}
where%
\begin{equation}
P_{21}^{n,m}=\left[  \Gamma_{21}+i\Delta+i\left(  n\delta+mw\right)  \right]
^{-1}, \label{sigma21mn2P}%
\end{equation}%
\begin{equation}
P_{12}^{n,m}=\left[  \Gamma_{21}-i\Delta+i\left(  n\delta+mw\right)  \right]
^{-1}, \label{sigma12mn2P}%
\end{equation}%
\begin{equation}
P_{m}^{n,m}=\frac{\gamma_{o}}{2}\left[  \gamma_{m}+\frac{\gamma_{o}}%
{2}+i\left(  n\delta+mw\right)  \right]  ^{-1}.
\end{equation}
We now solve these equations by writing an equation for the function
$\sigma_{0}^{m,n}$ alone. Substituting corresponding terms from Eqs.
(\ref{sigma21mn2}), (\ref{sigma12mn2}), and (\ref{sigma_m_mn}) into Eq.
(\ref{sigma0mn}), we obtain a two-dimensional recurrence relation for
$\sigma_{0}^{m,n}$%
\begin{align}
&  a_{n,m}^{1}\sigma_{0}^{n,m}+a_{n,m}^{2}\sigma_{0}^{n-2,m}+a_{n,m}^{3}%
\sigma_{0}^{n+2,m}+a_{n,m}^{4}\sigma_{0}^{n-1,m-1}+\nonumber\\
&  a_{n,m}^{5}\sigma_{0}^{n+1,m-1}+a_{n,m}^{6}\sigma_{0}^{n-1,m+1}+a_{n,m}%
^{7}\sigma_{0}^{n+1,m+1}+\nonumber\\
&  a_{n,m}^{8}\sigma_{0}^{n,m+2}+a_{n,m}^{9}\sigma_{0}^{n,m-2}=-R_{0,0}%
\delta_{n=0,m=0}, \label{RecRel}%
\end{align}
where we have introduced the following notation:%
\begin{align}
&  a_{n,m}^{1}=2\left\vert V_{21}^{s1}\right\vert ^{2}\left(  P_{21}%
^{n-1,m}+P_{12}^{n+1,m}\right)  +\nonumber\\
&  2\left\vert V_{21}^{s2}\right\vert ^{2}\left(  P_{21}^{n+1,m}%
+P_{12}^{n-1,m}\right)  +\nonumber\\
&  2\left\vert V_{21}^{p1}\right\vert ^{2}\left(  P_{21}^{n,m-1}%
+P_{12}^{n,m+1}\right)  +\nonumber\\
&  2\left\vert V_{21}^{p2}\right\vert ^{2}\left(  P_{21}^{n,m+1}%
+P_{12}^{n,m-1}\right)  +R_{n,m}, \label{F_A}%
\end{align}%
\begin{equation}
a_{n,m}^{2}=2V_{12}^{s1}V_{21}^{s2}\left(  P_{21}^{n-1,m}+P_{12}%
^{n-1,m}\right)  , \label{F_K}%
\end{equation}%
\begin{equation}
a_{n,m}^{3}=2V_{21}^{s1}V_{12}^{s2}\left(  P_{21}^{n+1,m}+P_{12}%
^{n+1,m}\right)  , \label{F_J}%
\end{equation}%
\begin{align}
a_{n,m}^{4}  &  =2V_{21}^{s2}V_{12}^{p1}\left(  P_{21}^{n,m-1}+P_{12}%
^{n-1,m}\right)  +\nonumber\\
&  2V_{12}^{s1}V_{21}^{p2}\left(  P_{21}^{n-1,m}+P_{12}^{n,m-1}\right)  ,
\label{F_I}%
\end{align}%
\begin{align}
a_{n,m}^{5}  &  =2V_{21}^{s1}V_{12}^{p1}\left(  P_{21}^{n,m-1}+P_{12}%
^{n+1,m}\right)  +\nonumber\\
&  2V_{12}^{s2}V_{21}^{p2}\left(  P_{21}^{n+1,m}+P_{12}^{n,m-1}\right)  ,
\label{F_F}%
\end{align}%
\begin{align}
a_{n,m}^{6}  &  =2V_{12}^{s1}V_{21}^{p}\left(  P_{21}^{n-1,m}+P_{12}%
^{n,m+1}\right)  +\nonumber\\
&  2V_{21}^{s2}V_{12}^{p2}\left(  P_{21}^{n,m+1}+P_{12}^{n-1,m}\right)  ,
\label{F_G}%
\end{align}%
\begin{align}
a_{n,m}^{7}  &  =2V_{12}^{s2}V_{21}^{p1}\left(  P_{21}^{n+1,m}+P_{12}%
^{n,m+1}\right)  +\nonumber\\
&  2V_{21}^{s1}V_{12}^{p2}\left(  P_{21}^{n,m+1}+P_{12}^{n+1,m}\right)  ,
\label{F_D}%
\end{align}%
\begin{equation}
a_{n,m}^{8}=2V_{21}^{p1}V_{12}^{p2}\left(  P_{21}^{n,m+1}+P_{12}%
^{n,m+1}\right)  , \label{a8}%
\end{equation}%
\begin{equation}
a_{n,m}^{9}=2V_{12}^{p1}V_{21}^{p2}\left(  P_{21}^{n,m-1}+P_{12}%
^{n,m-1}\right)  , \label{a9}%
\end{equation}%
\begin{equation}
R_{n,m}=\gamma_{21}+\frac{\gamma_{o}}{2}+i\left(  n\delta+mw\right)
-P_{m}^{n,m}\left(  \gamma_{21}-\gamma_{m}+\frac{\gamma_{o}}{2}\right)  .
\label{F_R}%
\end{equation}
The recurrence relation of Eq. (\ref{RecRel}) is solved numerically and allows
us to treat probe fields of arbitrary strength.

\subsection{Polarization}

The medium polarization is determined by the non-diagonal matrix element%
\begin{align}
P\left(  t\right)   &  =\mu N\left(  \rho_{21}+c.c.\right) \nonumber\\
&  =\mu N\left(
%TCIMACRO{\dsum \limits_{n=-\infty}^{+\infty}}%
%BeginExpansion
{\displaystyle\sum\limits_{n=-\infty}^{+\infty}}
%EndExpansion%
%TCIMACRO{\dsum \limits_{m=-\infty}^{+\infty}}%
%BeginExpansion
{\displaystyle\sum\limits_{m=-\infty}^{+\infty}}
%EndExpansion
\sigma_{21}^{n,m}e^{-i\omega_{0}t}e^{i\left(  n\delta+mw\right)
t}+c.c.\right)  . \label{Polarization}%
\end{align}
Thus the component of the polarization that oscillates at a particular
frequency gives the medium's response to the electromagnetic field at this
frequency. As can be seen from Eq. (\ref{Polarization}), the polarization at
the probe frequency $\omega_{p1}$ is proportional to $\sigma_{21}^{0,-1}$
which can be found from Eq. (\ref{sigma21mn2})%
\begin{align}
\sigma_{21}^{0,-1}  &  =iP_{21}^{0,-1}\nonumber\\
&  \left(  V_{21}^{s1}\sigma_{0}^{1,-1}+V_{21}^{s2}\sigma_{0}^{-1,-1}%
+V_{21}^{p1}\sigma_{0}^{0,0}+V_{21}^{p2}\sigma_{0}^{0,-2}\right)  .
\label{ProbePolarization}%
\end{align}
In the next section we analyze the normalized probe absorption, which is
defined as $\alpha/\alpha_{0}=-\operatorname{Im}\sigma_{21}^{0,-1}\Gamma
_{21}/V_{21}^{p1}$, for several cases.

\section{\label{sec:level3}The CPO resonance in monochromatic pumping}

The role of the excited-state decay's branching ratio to the ground and
metastable states is not discussed in the literature. However, it is of
crucial importance in obtaining CPO at the subharmonic frequencies. Thus, in
order to better understand the behavior of the system in a bichromatic driving
field, we first consider the case of a monochromatic driving where it is easy
to get an analytical result and analyze the role of the branching ratio. Let
us consider the interaction of the TLS\ with a strong pump and a weak probe.
In this case, we take $V_{21}^{s2}=V_{21}^{p2}=0$ in Eq.
(\ref{ProbePolarization}) so that%
\begin{equation}
\sigma_{21}^{0,-1}=iP_{21}^{0,-1}\left(  V_{21}^{s1}\sigma_{0}^{1,-1}%
+V_{21}^{p1}\sigma_{0}^{0,0}\right)  .
\end{equation}
Assuming that the probe field is weak we can keep only the first-order terms
in the probe Rabi frequencies. We then obtain for the functions $\sigma
_{0}^{1,-1}$ and $\sigma_{0}^{0,0}$ of Eq. (\ref{RecRel})
\begin{equation}
\sigma_{0}^{1,-1}=-\frac{a_{1,-1}^{6}}{a_{1,-1}^{1}}\sigma_{0}^{0,0},
\label{sigma0m_probe}%
\end{equation}%
\begin{equation}
\sigma_{0}^{0,0}=-R_{0,0}/a_{0,0}^{1}=-\left(  1-\frac{\kappa}{1+\kappa
}\right)  , \label{SigmaZeroSingleDrSol}%
\end{equation}
where
\begin{equation}
\kappa=\kappa_{TLS}\left(  \gamma_{0}/2\gamma_{m}+1\right)
\label{saturationParameter}%
\end{equation}
is the saturation parameter of the system and%
\begin{equation}
\kappa_{TLS}=4\frac{\left\vert V_{12}^{1s}\right\vert ^{2}}{\gamma_{2}%
\Gamma_{21}}\frac{1}{1+\Delta_{s}^{2}/\Gamma_{21}^{2}}
\label{saturationParameterTLS}%
\end{equation}
is the saturation parameter of the ordinary two-level system (if there is no
additional channel for the decay via intermediate state). The functions
$a_{1,-1}^{1}$ and $a_{1,-1}^{6}$ are found from Eqs. (\ref{F_A}) and
(\ref{F_G}) to be%
\begin{align}
a_{1,-1}^{1}  &  =\frac{\left[  \gamma_{2}-i\Delta_{p}\right]  \left[
\gamma_{m}-i\Delta_{p}\right]  }{\gamma_{m}+\gamma_{o}/2-i\Delta_{p}%
}+\nonumber\\
&  4\left\vert V_{21}^{s1}\right\vert ^{2}\frac{\Gamma_{21}-i\Delta_{p}%
}{\left[  \Gamma_{21}+i\left(  \Delta_{s}+\Delta_{p}\right)  \right]  \left[
\Gamma_{21}-i\left(  \Delta_{s}-\Delta_{p}\right)  \right]  }, \label{A0m}%
\end{align}%
\begin{equation}
a_{1,-1}^{6}=4V_{12}^{0}V_{21}^{p1}\frac{\Gamma_{21}+i\left(  \Delta
_{s}-\Delta_{p}\right)  /2}{\left(  \Gamma_{21}+i\Delta_{s}\right)  \left(
\Gamma_{21}-i\Delta_{p}\right)  }, \label{B0m}%
\end{equation}
where for convenience we have introduced the pump detuning from resonance
$\Delta_{s}\equiv\Delta-\delta=\omega_{s1}-\omega_{21}$ and pump-probe
detuning $\Delta_{p}\equiv w-\delta=\omega_{p1}-\omega_{s1}$.

We are interested in the shape of the narrow dip centered at the pump
frequency and associated with the CPO effect. This structure has a
characteristic bandwidth of the order of $\gamma_{m}$ which is much less than
the natural bandwidth of the transition from the excited to the ground state
as the level $\left\vert m\right\rangle $ is assumed to be metastable. Thus,
in Eqs. (\ref{A0m}) and (\ref{B0m}), we use the approximation%
\begin{equation}
\gamma_{m},\Delta_{p}\ll\Gamma_{21},\gamma_{2}%
\end{equation}
and obtain%
\begin{equation}
\frac{a_{1,-1}^{6}}{a_{1,-1}^{1}}\approx4\frac{V_{12}^{s1}V_{21}^{p1}}%
{\gamma_{2}\Gamma_{21}\left(  1+\Delta_{s}^{2}/\Gamma_{21}^{2}\right)  }%
\frac{S}{1+\kappa_{TLS}}\frac{\gamma_{m}+\gamma_{0}/2-i\Delta_{p}}{\left(
W-i\Delta_{p}\right)  }, \label{ABappr}%
\end{equation}
where $W$ is the bandwidth of the CPO dip%
\begin{equation}
W=\gamma_{m}\frac{1+\kappa}{1+\kappa_{TLS}}. \label{CPO_bandwidth}%
\end{equation}
and the parameter $S\equiv\left(  1+\Delta_{s}^{2}/2\Gamma_{21}^{2}%
-i\Delta_{s}/2\Gamma_{21}\right)  $. Combining Eqs. (\ref{sigma0m_probe}),
(\ref{SigmaZeroSingleDrSol}), and (\ref{ABappr}), we obtain the medium
response at the probe frequency in the region of the CPO resonance%
\begin{equation}
\sigma_{21}^{0,-1}=\frac{-iV_{21}^{p1}}{1+\kappa}\frac{1}{\Gamma_{21}+i\left(
\Delta_{s}-\Delta_{p}\right)  }\left(  1-XS\right)  ,
\label{sigma0m_probeSolution}%
\end{equation}
where the first term in the brackets determines the saturation of the system
due to the pump and the second term $X$ includes the coherent interaction
between pump and probe fields%
\begin{equation}
X=1-\frac{1}{1+\kappa_{TLS}}\frac{\gamma_{m}-i\Delta_{p}\kappa_{TLS}%
}{W-i\Delta_{p}}. \label{CoherentInteractionTerm}%
\end{equation}
Further, for simplicity we consider the case of small pump detunings and
assume that $S\approx1$.

\textbf{Bandwidth} The bandwidth of the resonance is determined by the
function $W$ [see Eq. (\ref{CPO_bandwidth})]. In the case of the ``classical"
CPO, for which $\gamma_{2}\approx\gamma_{o}\gg\gamma_{m}$, we have $\kappa
\gg\kappa_{TLS}$, so that in the denominator of Eq. (\ref{CPO_bandwidth}) we
can omit $\kappa_{TLS}$ and obtain the same result as in
\cite{SharypovPhaseDep}%

\begin{equation}
W=\gamma_{m}\left(  1+\kappa\right)  . \label{BandWidthHighCPO}%
\end{equation}
Thus the minimal bandwidth of the CPO resonance is determined by $\gamma_{m}$
and this resonance experiences power broadening. In the opposite case for
which $\gamma_{2}\gg\gamma_{o}\approx\gamma_{m}$, $\kappa$ and $\kappa_{TLS}$
are of the same order. In this case, the bandwidth experiences fast saturation%
\begin{equation}
W=\gamma_{m}+\frac{\gamma_{o}}{2}\frac{\kappa_{TLS}}{1+\kappa_{TLS}}
\label{BandWidthLowCPO}%
\end{equation}
and will not exceed $\gamma_{m}+\frac{\gamma_{o}}{2}$.

\textbf{Depth} The depth is determined from Eq. (\ref{sigma0m_probeSolution})
by putting $\Delta_{p}=0$. We obtain $\alpha\left(  \Delta_{p}=0\right)
/\alpha_{0}=1/{\left(  1+\kappa\right)  ^{2}}$.

\begin{figure}
[ptb]
\begin{center}
\includegraphics[
height=1.497in,
width=2.0678in
]%
{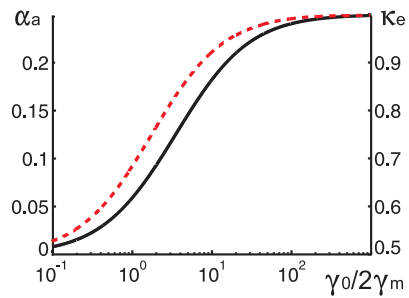}%
\caption{The extremum point $\kappa_{e}$ (black line) and dip amplitude at
this poin $\alpha_{a}\left(  \kappa_{e}\right)  $ (red line) as a function of
$\gamma_{o}/2\gamma_{m}$. }%
\label{SparamNUM}%
\end{center}
\end{figure}

\textbf{Amplitude} From Eq. (\ref{sigma0m_probeSolution}), we can determine
the amplitude of the CPO dip. The dip itself comes from the multiplier
$({\gamma_{m}-i\Delta_{p}}\kappa_{TLS})/({W-i\Delta_{p}})$ in Eq.
(\ref{CoherentInteractionTerm}). Its amplitude $\alpha_{a}$ is given by the
difference between Eq. (\ref{sigma0m_probeSolution}) with and without this
term at point $\Delta_{p}=0$%
\begin{equation}
\alpha_{a}=\frac{\kappa}{\left(  1+\kappa\right)  ^{2}}\frac{\gamma
_{o}/2\gamma_{m}}{\gamma_{o}/2\gamma_{m}+1+\kappa}. \label{DipAmplitude}%
\end{equation}
It follows from Eq. (\ref{DipAmplitude}) that, in the high saturation regime,
the amplitude of the dip decreases as $1/|V_{21}^{s1}|^{4}$. The function
$\alpha_{a}$ has an extremum at the point%
\begin{equation}
\kappa_{e}=\frac{\gamma_{0}/2\gamma_{m}+1}{4}\left(  \sqrt{1+8\left(
\gamma_{0}/2\gamma_{m}+1\right)  ^{-1}}-1\right)  . \label{CPOamplitudeMAX}%
\end{equation}
For the case in which $\gamma_{o}\gg\gamma_{m}$ it is easy to show that the
maximum amplitude of the dip $\alpha_{a}=0.25$ is obtained at the point
$\kappa_{e}=1$ \cite{SharypovPhaseDep}, but in the general situation there is
no simple expression so that a graphical representation is required (see Fig.
\ref{SparamNUM}).%
%TCIMACRO{\FRAME{ftbpFU}{2.0678in}{1.497in}{0pt}{\Qcb{The extremum point
%$\kappa_{e}$ (black line) and dip amplitude at this poin $\alpha_{a}\left(
%\kappa_{e}\right)  $ (red line) as a function of $\gamma_{o}/2\gamma_{m}$. }%
%}{\Qlb{SparamNUM}}{spar.eps}{\special{ language "Scientific Word";
%type "GRAPHIC";  maintain-aspect-ratio TRUE;  display "USEDEF";
%valid_file "F";  width 2.0678in;  height 1.497in;  depth 0pt;
%original-width 5.0981in;  original-height 3.6746in;  cropleft "0";
%croptop "1";  cropright "1";  cropbottom "0";
%filename 'Spar.eps';file-properties "XNPEU";}} }%
%BeginExpansion
%EndExpansion
It can be seen from Fig. \ref{SparamNUM} that with decreasing $\gamma
_{o}/\gamma_{m}$ the maximal achievable dip amplitude decreases.

\section{\label{sec:level4}Numerical results}

In this section we numerically analyze the absorption spectrum of a weak probe
field at the frequency $\omega_{p1}$ that scans the TLS (Fig.\ref{IntScheme}).
The TLS is driven by two pump fields that are symmetrically detuned from
resonance ($\Delta=0$). Also we consider the case when there can be an
additional weak field whose detuning is symmetrically displaced from resonance
with respect to the first probe [see Eq. (\ref{freq_notation})]. In this case,
we demonstrate strong parametric interaction between the weak fields in the
presence of the pumps. As we mentioned in the previous sections, the branching
ratio of $\gamma_{21}$, the decay rate from the excited to the ground state,
to $\gamma_{o}$, the decay rate from the excited to the metastable state,
plays an important role in the properties of the CPO. We therefore analyze two
cases numerically: $\gamma_{2}\approx\gamma_{o}\gg\gamma_{m}$ and $\gamma
_{2}\gg\gamma_{o}\approx\gamma_{m}$.

\subsection{Case 1 $\gamma_{2}\approx\gamma_{o}\gg\gamma_{m}$}

\begin{figure}
[ptb]
\begin{center}
\includegraphics[
height=1.214in,
width=3.4786in
]%
{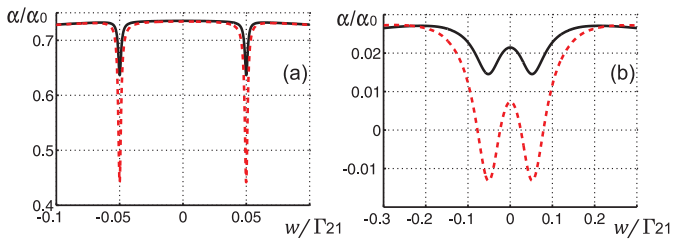}%
\caption{The spectrum of the probe absorption as the function of the detuning
$w$. $\Delta=0$, $\delta/\Gamma_{21}=0.05$, $\gamma_{o}/\gamma_{2}=0.9$,
$\gamma_{21}/\gamma_{2}=0.1$, $\gamma_{m}/\Gamma_{21}=10^{-3}$, $\gamma
_{2}/\Gamma_{21}=2,~V^{p1}\ll\Gamma_{21}$. a) $V^{s1}/\Gamma_{21}%
=V^{s2}/\Gamma_{21}=0.01$ ($\kappa=0.36$). b) $V^{s1}/\Gamma_{21}%
=V^{s2}/\Gamma_{21}=0.1$ ($\kappa=34$). $V^{p2}=0$ black line. $V^{p2}%
/V^{p1}=2$ red line.}%
\label{HCPO}%
\end{center}
\end{figure}

This range of values for the relaxation constants has been considered in our
previous work on CPO \cite{SharypovPhaseDep,IdoCPO,AsiCPO,HellerSharypov}.
However, here two strong pump fields are applied to the same transition. So,
instead one CPO resonance in the probe spectrum, we obtain two CPO dips
centered at the pump frequencies $w=\pm\delta$\ (Fig. \ref{HCPO}). According
to Eq. (\ref{BandWidthHighCPO}), increasing the pump Rabi frequencies leads to
considerable power broadening of the CPO resonances [Fig. \ref{HCPO}(b)]. The
bandwidth of the CPO resonances estimated from the numerical simulation is
well described by Eq. (\ref{BandWidthHighCPO}) which was derived for the case
in which there is only one strong field \cite{N1}. The behavior of each of
these two resonances appears to be independent - while the probe field is in
the region of the CPO resonance created by the first pump, it does not
experience any coherent effects that come from the second pump field. The
situation became quite different when we add a second symmetrically displaced
weak field [Fig. \ref{HCPO} (red dashed lines)].%
%TCIMACRO{\FRAME{ftbpFU}{3.4786in}{1.214in}{0pt}{\Qcb{The spectrum of the probe
%absorption as the function of the detuning $w$. $\Delta=0$, $\delta
%/\Gamma_{21}=0.05$, $\gamma_{o}/\gamma_{2}=0.9$, $\gamma_{21}/\gamma_{2}=0.1$,
%$\gamma_{m}/\Gamma_{21}=10^{-3}$, $\gamma_{2}/\Gamma_{21}=2,~V^{p1}\ll
%\Gamma_{21}$. a) $V^{s1}/\Gamma_{21}=V^{s2}/\Gamma_{21}=0.01$ ($\kappa=0.36$).
%b) $V^{s1}/\Gamma_{21}=V^{s2}/\Gamma_{21}=0.1$ ($\kappa=34$). $V^{p2}=0$ black
%line. $V^{p2}/V^{p1}=2$ red line.}}{\Qlb{HCPO}}{hcpo.eps}%
%{\special{ language "Scientific Word";  type "GRAPHIC";
%maintain-aspect-ratio TRUE;  display "USEDEF";  valid_file "F";
%width 3.4786in;  height 1.214in;  depth 0pt;  original-width 16.4946in;
%original-height 7.9212in;  cropleft "0";  croptop "1";  cropright "1";
%cropbottom "0";  filename 'HCPO.eps';file-properties "XNPEU";}} }%
%BeginExpansion
%EndExpansion
In the presence of the second weak field, FWM occurs. Due to the effect of the
FWM, the absorption of two photons from the pump fields at the frequencies
$\omega_{s1}$ and $\omega_{s2}$ leads to the generation of two photons at the
probe frequencies $\omega_{p1}$ and $\omega_{p2}$. As a result, the dips in
Fig. \ref{HCPO}(a) get deeper and there is even amplification of the probe in
Fig. \ref{HCPO}(b).

The effect of FWM between two weak probes and two pump fields exists (see red
dashed lines in Fig. \ref{HCPO}) even when the pump Rabi frequencies are well
below the transverse relaxation rate $\Gamma_{21}$. The amplitudes of the two
probes in Fig. \ref{HCPO} are chosen to be unequal in order to obtain a more
pronounced effect than in the spectrum of a single probe
\cite{SharypovPhaseDep}.

In the present case there are no CPO resonances at the subharmonic frequencies
$w=\pm\left(  2n+ 1\right)  \delta$. This is due to the large difference in
the saturation parameters for the TLS in the presence and absence of the
intermediate state, as shown in Eq. (\ref{saturationParameter}). Thus, when
the Rabi frequencies are close to the transverse relaxation rate of the TLS
where the higher-order nonlinear wave mixing appears \cite{Ficek1993}, the CPO
resonances get very broad [see Eq. (\ref{BandWidthHighCPO})] and completely disappear.

\subsection{Case 2 $\gamma_{2}\gg\gamma_{o}\approx\gamma_{m}$}

\begin{figure}
[ptb]
\begin{center}
\includegraphics[
height=2.4132in,
width=3.4067in
]%
{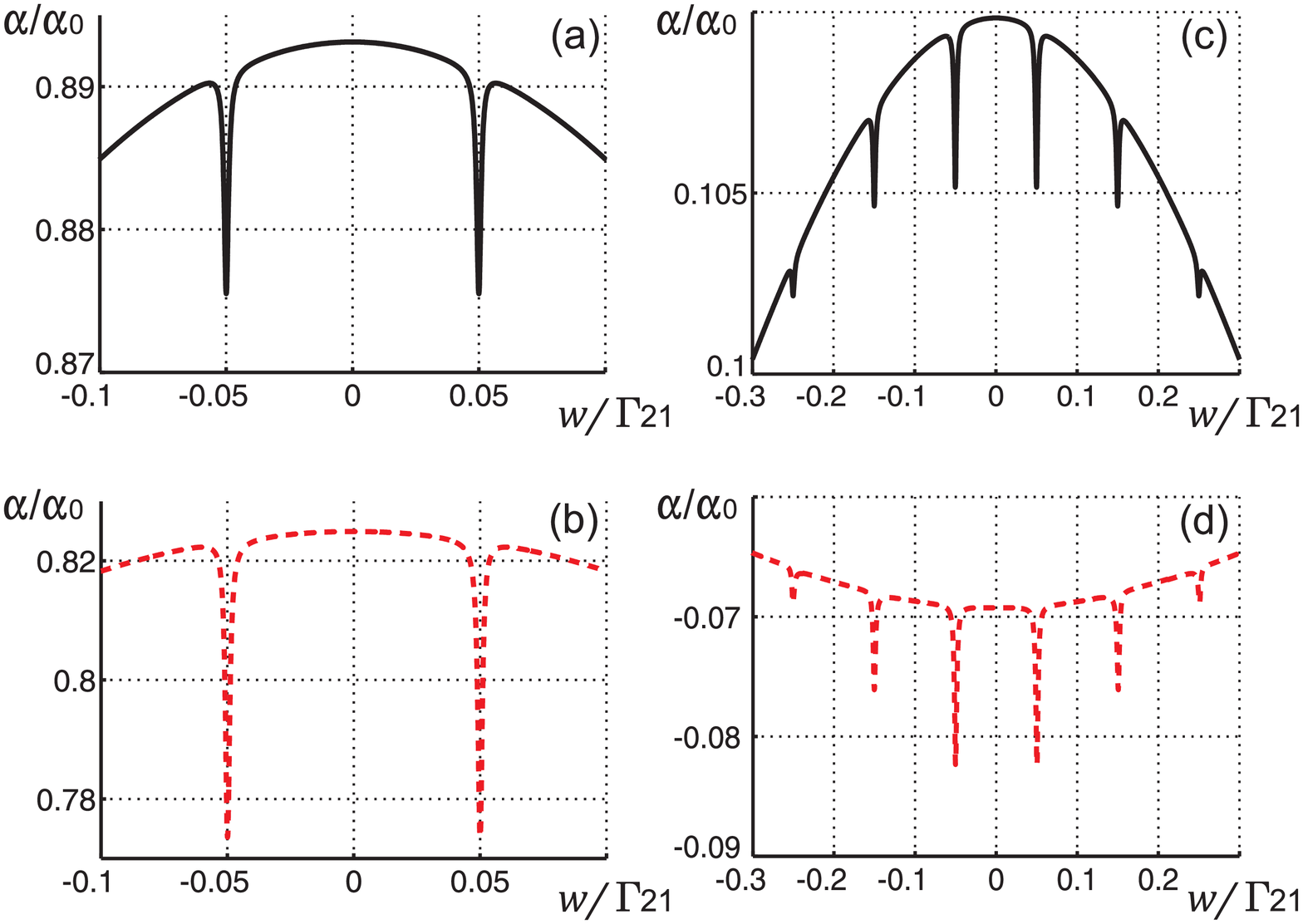}%
\caption{The spectrum of the probe absorption as the function of the detuning
$w$. $\Delta=0$, $\delta/\Gamma_{21}=0.05$, $\gamma_{o}/\gamma_{2}=10^{-3}$,
$\gamma_{21}/\gamma_{2}=0.999$, $\gamma_{m}/\Gamma_{21}=10^{-3}$, $\gamma
_{2}/\Gamma_{21}=2,~V^{p1}\ll\Gamma_{21}$. When the pump Rabi frequensies are
much smaller than a transverse relaxation rate $V^{s1}/\Gamma_{21}%
=V^{s2}/\Gamma_{21}=0.1$ ($\kappa=0.08$) there are only two CPO resonances
a,b) and the resonanses at the subharmonics appears when the pump Rabi
frequensies are increased $V^{s1}/\Gamma_{21}=V^{s2}/\Gamma_{21}=1$
($\kappa=4$) c,d). The presence of the second probe $V^{p2}/V^{p1}%
=2$\ demonstrate the parametric interaction between the weak fields b,d)
($V^{p2}=0$ for a,c)).}%
\label{LCPO}%
\end{center}
\end{figure}

In this range of relaxation parameters, almost all the population decays
directly from the excited to the ground state and only a tiny amount decays
via the metastable state. In this case, the two CPO resonances centered at the
pump frequencies are still observed but their amplitudes get much smaller
(compare Figs. \ref{HCPO} and \ref{LCPO}). The special feature in this case is
that $\kappa$ and $\kappa_{TLS}$\ are of the same order [see Eq.
(\ref{saturationParameter})] so that the CPO resonances do not experience much
power broadening [see Eq. (\ref{BandWidthLowCPO})]. These facts allow one to
combine the effect of the higher-order wave mixing which appears at pump Rabi
frequencies comparable with the transverse relaxation rate $\Gamma_{21}$
\cite{Ficek1993}\ and the existence of narrowband CPO resonances. In Figs.
\ref{LCPO}(c) and (d), the existence of the CPO resonances at the subharmonic
frequencies $w=\pm\left(  2n+1\right)  \delta$\ where $n=0,1,2,3...$ is
clearly seen. For example the resonance that is located at the $w/\Gamma
_{21}=0.15$ ($n=1$) on the Fig. \ref{LCPO}(c) corresponds to the process of
the three-wave mixing between pump fields $\omega_{sub\_0.15}=2\omega
_{s1}-\omega_{s2}$.%
%TCIMACRO{\FRAME{ftbpFU}{3.4067in}{2.4132in}{0pt}{\Qcb{The spectrum of the
%probe absorption as the function of the detuning $w$. $\Delta=0$,
%$\delta/\Gamma_{21}=0.05$, $\gamma_{o}/\gamma_{2}=10^{-3}$, $\gamma
%_{21}/\gamma_{2}=0.999$, $\gamma_{m}/\Gamma_{21}=10^{-3}$, $\gamma_{2}%
%/\Gamma_{21}=2,~V^{p1}\ll\Gamma_{21}$. When the pump Rabi frequensies are much
%smaller than a transverse relaxation rate $V^{s1}/\Gamma_{21}=V^{s2}%
%/\Gamma_{21}=0.1$ ($\kappa=0.08$) there are only two CPO resonances a,b) and
%the resonanses at the subharmonics appears when the pump Rabi frequensies are
%increased $V^{s1}/\Gamma_{21}=V^{s2}/\Gamma_{21}=1$ ($\kappa=4$) c,d). The
%presence of the second probe $V^{p2}/V^{p1}=2$\ demonstrate the parametric
%interaction between the weak fields b,d) ($V^{p2}=0$ for a,c)).}}{\Qlb{LCPO}%
%}{lcpo.eps}{\special{ language "Scientific Word";  type "GRAPHIC";
%maintain-aspect-ratio TRUE;  display "USEDEF";  valid_file "F";
%width 3.4067in;  height 2.4132in;  depth 0pt;  original-width 17.2511in;
%original-height 8.1749in;  cropleft "0";  croptop "1";  cropright "1";
%cropbottom "0";  filename 'LCPO.eps';file-properties "XNPEU";}} }%
%BeginExpansion
%EndExpansion
In this case, as in the previous one, there is strong parametric interaction
between the weak symmetrically displaced fields [Fig. \ref{LCPO}(b) and (d)].
In addition, the pump Rabi frequencies are of the same order as the transverse
relaxation rate. This leads to parametric interaction between the weak fields
not only in the CPO region, as in the previous case, but in a broad spectral
range, as shown in Fig. \ref{LCPO}(d) \cite{Boyd1981}.

\section{\label{sec:level5}Conclusion}

In summary, we have demonstrated the appearance of narrowband CPO responses at
frequencies shifted from the pump frequency. The new CPO resonances appears at
the frequencies $\omega_{sub}=\omega_{0}\pm\left(  2n+1\right)  \delta$ due to
multi-photon mixing between components of the bichromatic pump. We expect that
these spectral features will allow spectral filtration of the signal field
from the strong pump in experiments with slow light, and biphoton generation
based on the effect of CPO. Also, in the region of the CPO dip there is
effective parametric interaction between probe fields for both the cases
considered in Sec. IV. The subharmonic resonances that appear when the
condition $\gamma_{2}\gg\gamma_{o}\approx\gamma_{m}$ is satisfied can be
investigated in atomic TLS systems where the decay via dipole-forbidden
transitions are taken into account, and in certain NV centers in diamond
\cite{NVdiamond}. In addition, the predicted subharmonic CPO resonances may
have contributions to the resonances observed in the recent experiment
\cite{Akulshin2011} where the was driven by a bichromatic pump.

\acknowledgements A. V. S. was supported by RFBR 12-02-31621


\begin{thebibliography}{99}
\bibitem {N1}The values of $\kappa$ given in the captions of Figs. \ref{HCPO}
and \ref{LCPO} are found under the assumption that the population inversion
can always be written in the form of Eq. (\ref{SigmaZeroSingleDrSol}) where
$\sigma_{0}^{0,0}$ (zero-order population inversion)\ is found from the
numerical solution.
\end{thebibliography}
\end{document}